\documentclass[%
 reprint,
nofootinbib,
 amsmath,amssymb,
 aps,
floatfix,
]{revtex4-2}

\usepackage{placeins}

\usepackage{graphicx}%
\usepackage{dcolumn}%
\usepackage{bm}%
\usepackage{amsmath}
\usepackage{amsfonts}
\usepackage{subcaption}
\usepackage{physics}
\usepackage{pifont} %
\newcommand{\cmark}{\ding{51}} %
\newcommand{\xmark}{\ding{55}} %

\usepackage{makecell}  %
\usepackage{tcolorbox} %
\usepackage{tikz} %

\newcommand{\Ts}{T_\text{S}}
\newcommand{\Exc}{E_\text{XC}}
\newcommand{\Etxc}{E_\text{TXC}}
\newcommand{\EH}{E_\text{H}}
\newcommand{\Eext}{E_\text{ext}}
\newcommand{\Etot}{E_\text{tot}}
\newcommand{\Egs}{E^*}

\newcommand{\ES}{\tilde{E}} %
\newcommand{\EStot}{\tilde{E}_\text{tot}} %
\newcommand{\EStxc}{\tilde{E}_\text{TXC}} %
\newcommand{\Eaux}{\tilde{E}_\text{aux}}
\newcommand{\EWS}{\bar{E}} %

\newcommand{\coulomb}{\mathbf{J}}

\newcommand{\ovlp}{\mathbf{S}}

\newcommand{\vp}{\mathbf{p}}
\newcommand{\vpgs}{\mathbf{p}^*}
\newcommand{\dens}{\rho}
\newcommand{\densgs}{\rho^*} %

\newcommand{\vecr}{\mathbf{r}}

\newcommand{\argmin}{\arg\min}

\begin{document}

\preprint{APS/123-QED}

\title{Orbital-Free Surrogate Functionals\\Yield Transferable Interatomic Potentials and Electron Densities}

\author{Simon Wagner}
\author{Marc K. Ickler}%
\author{Manuel V. Klockow}
\author{Fred A. Hamprecht}
\author{Roman Remme}
\email{Corresponding author: roman.remme@iwr.uni-heidelberg.de}
\affiliation{Interdisciplinary Center for Scientific Computing (IWR), Heidelberg University,
Heidelberg, 
Germany}%

\date{\today}%

\begin{abstract}
Orbital-free density functional theory seeks to compute the energy of an electronic system directly from its electron density, avoiding one-electron wave functions and thereby offering a route to scalable electronic structure calculations.
Machine-learned orbital-free density functionals have recently achieved promising results on small organic molecules, predicting energies with sub-millihartree accuracy. However, their convergence in density optimization remains sensitive to hyperparameter tuning and architectural choices. Here, we extend the recently introduced (weak) surrogate functional framework---designed to predict ground-state electron densities only---to also yield their energy, resulting in ``strong'' surrogate functionals. We find that these learned functionals enable stable convergence across all tested neural network backbones, reducing electron density errors relative to the Kohn--Sham reference by an order of magnitude compared to previous OF-DFT methods. More importantly, the predicted energies are competitive with state-of-the-art machine-learned interatomic potentials (MLIPs) trained only on energies, while exhibiting superior generalization to larger, unseen systems.
\end{abstract}

\maketitle

\section{\label{sec:introduction} Introduction}

Kohn--Sham density functional theory (KS-DFT)~\cite{kohn1965self} is, for good reasons, a mainstay of quantum chemistry, but the cubic scaling of its standard formulation limits the sizes of systems and lengths of dynamics that can be studied at an acceptable computational cost.
Orbital-free DFT (OF-DFT) sidesteps the explicit treatment of orbitals and minimizes the electronic energy directly with respect to the density~\cite{wang1999orbital, mi2023orbital}, promising a more favorable scaling.
Realizing this promise in practice, however, requires
an accurate approximation of the energy functional, in particular of its kinetic energy contribution~\cite{mi2023orbital}.

Machine learning has recently begun to address this challenge~\cite{chen2026machine}, with progress accelerating from pioneering work on one-dimensional toy systems~\cite{snyder2013orbital, meyer2020machine} through small molecules~\cite{yao2016kinetic, seino2018semi-local, fujinami2020orbital-free, golub2019kinetic, remme2023kineticnet} to mid-sized molecules~\cite{zhang2024overcoming, remme2025stable}.
Among these, M-OFDFT~\cite{zhang2024overcoming} was the first to predict molecular energies with errors on the order of 1 mHa compared to the Kohn--Sham reference across the QM9 dataset~\cite{ramakrishnan2014quantum}, though the learned functional did not afford true variational optimization.
STRUCTURES25~\cite{remme2025stable} closed this gap by augmenting the training data with densities obtained from perturbed Kohn--Sham potentials, yielding a functional that converges from a cheap initial guess to a meaningful ground-state density. 
Most recently, weak \textit{surrogate functionals}~\cite{remme2026surrogate} have shown that such perturbations are not even strictly necessary: a learned functional that is only required to yield the correct density upon minimization at inference time can be trained from in-vacuo ground-state labels alone. 

Despite this progress, two limitations persist.
First, by design, weak surrogates predict only the ground-state \emph{density}; the energy at the converged density is not constrained and is therefore not physically meaningful. 
A practical OF-DFT functional, however, must deliver both: a reliable density for downstream electronic structure analysis (e.g., electrostatics, response properties, bonding) and an energy
to drive geometry optimization and molecular dynamics.
Second, even existing convergent OF-DFT functionals have proven sensitive to architectural and hyperparameter choices: minor modifications can turn an otherwise well-behaved energy surface into one lacking a usable minimum, or one whose minimum lies far from the true ground state. 
This fragility hampers the systematic exploration of more expressive neural network backbones and slows the transfer of the approach to new chemistries.

\begin{figure*}[ht]
    \centering
    \hfill
    \begin{subfigure}{0.4\textwidth}
        \centering
        \includegraphics[width=\linewidth]{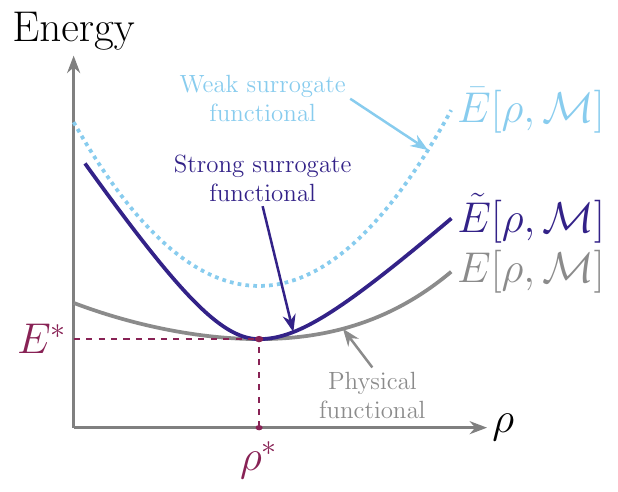}
        \caption{}
        \label{fig:functional}
    \end{subfigure}
    \hfill
    \begin{subfigure}{0.5\textwidth}
        \centering
        \includegraphics[width=\linewidth]{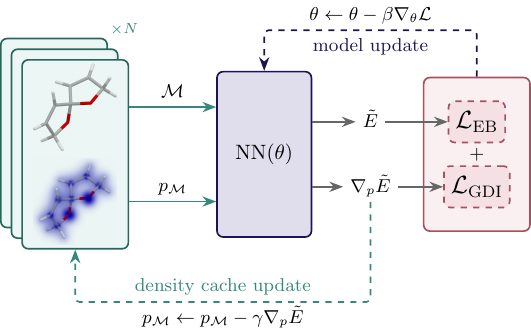}
        \caption{}
        \label{fig:pipeline}
    \end{subfigure}
    \hfill
    \caption{Overview of the strong surrogate pipeline. (a) The strong surrogate functional $\ES$ is required to have a global minimum at the true ground-state density $\densgs$ with energy equal to the true ground-state energy $\Egs$. In contrast, a weak surrogate functional $\EWS$~\cite{remme2026surrogate} is only required to have a minimum at $\densgs$, while the energy is not constrained. (b) Simplified overview of the train-time density optimization procedure. For each molecule, we cache its current density. In each training step, we update the densities of the involved molecules with a gradient-descent step using the gradients predicted by the model.}
    \label{fig:surrogate_overview}
\end{figure*}

Here, we introduce \textit{strong surrogate functionals}, which extend the weak surrogate framework~\cite{remme2026surrogate} to constrain not only the location of the minimum but also its energy.
As illustrated in Figure~\ref{fig:functional}, a strong surrogate is required to possess a global minimum at the true ground-state density \emph{and} to match the true ground-state energy at that minimum.
Constructing such a functional from ground-state labels alone presents the challenge to provide a training signal for off-equilibrium energies where the ground-truth is not known.
We address this by combining two core ingredients: (1) ground-state energy and gradient supervision by injecting a number of explicit ground-state samples per batch to anchor the absolute energy scale, and (2) an energy-bound loss derived from empirical parabolic envelopes of the Kohn--Sham energy landscape.
Together, these components allow the model to inherit the robust convergence properties of weak surrogates~\cite{remme2026surrogate} while simultaneously yielding accurate ground-state energies. %

Empirically, strong surrogates deliver the following improvements over the prior state of the art:
\begin{enumerate}
    \item Density optimization \emph{converges across all neural-network backbones we tested}, including eSEN~\cite{fu2025learning} and EquiformerV2~\cite{liao2024equiformerv2}.
    The fragility that has so far limited the choice of architecture in machine-learned OF-DFT is largely removed.
    \item Density errors on QM9 drop by an order of magnitude relative to STRUCTURES25 and weak surrogates, and similar improvements carry over to QMugs molecules with up to 100 heavy atoms~\cite{isert2022qmugs}---well beyond the size range seen during training.
    \item The predicted energies are competitive with current state-of-the-art MLIPs~\cite{fu2025learning, aykent2025gotennet} that are trained directly on energies, while extrapolating substantially better to larger systems---closing the gap between density-based and geometry-based machine learning models for molecules.
    \item The framework naturally accommodates training on partial functional components, such as the kinetic and exchange-correlation energy ($\Etxc$), allowing exact analytical computation of nonlocal electrostatic contributions while maintaining stable density optimization.
\end{enumerate}

\section{\label{sec:related_work} Related work}

\paragraph{Machine-learned orbital-free DFT.}
Recent advances in machine-learned OF-DFT~\cite{zhang2024overcoming, remme2025stable} have focused on learning a density functional using the densities and energies of intermediate Kohn--Sham iterations as ground-truth labels, allowing the prediction of the electronic ground state through variational optimization. Weak surrogate functionals~\cite{remme2026surrogate} bypass the requirement of physical off-equilibrium labels by dropping physical constraints and only requiring the learned functional to have a global minimum at the true ground-state density. At the same time, they lose the ability to predict meaningful energies, a property that is reintroduced in the strong surrogate functional framework presented here.

\paragraph{Machine-learned interatomic potentials (MLIPs).}
In contrast to density-based methods, standard MLIPs~\cite{schutt2018schnet, batzner2022nequip, batatia2022mace, musaelian2023learning, liao2024equiformerv2, fu2025learning, aykent2025gotennet} bypass the electron density entirely to efficiently predict the potential energy surface directly from the molecular geometry. While modern equivariant graph neural networks achieve outstanding interpolation accuracy, the lack of an explicit intermediate density restricts their ability to natively describe, e.g., nonlocal electronic responses and charge transfer.

\paragraph{Long-range and self-consistent electrostatic MLIPs.}
To overcome the limitations of strictly local MLIPs, several approaches augment
geometry-based models with coarse-grained electronic degrees of freedom or
explicitly long-ranged descriptors~\cite{anstine2023machine}. One family has
been systematized as a design space of self-consistent electrostatic MLIPs,
interpretable as coarse-grained approximations to DFT~\cite{baldwin2026design}.
Examples include charge-density or charge-equilibration
schemes~\cite{ko2021fourth}, self-consistent electronic
populations~\cite{xie2020incorporating}, long-range polarization
models~\cite{gao2022self}, and explicit-electron
representations~\cite{cools2022modeling}.
A complementary line of work captures long-range interactions without
imposing electronic self-consistency, e.g., through physics-inspired
long-range descriptors~\cite{grisafi2019incorporating}, directly-predicted
Wannier-center representations~\cite{zhang2022deep}, Ewald-summation-based
descriptors and message passing~\cite{cheng2025latent, kosmala2023ewald}, global
attention mechanisms~\cite{frank2026euclidean}, or charge-aware potentials and
electric-response models with explicit electronic degrees of
freedom~\cite{isayev2025aimnet2, falletta2025unified}.
Common to all these approaches is that the electronic degrees of freedom, where present, need not coincide with the true DFT ground state. Strong surrogate functionals, in contrast, are able to capture the true DFT ground-state density at a much higher level of detail.
\section{Background}

\subsection{Orbital-free density functional theory}

In OF-DFT, the goal is to construct a functional $E[\dens]$, which for a given electronic system maps an electron density $\dens$ to a corresponding energy $E$. Minimizing $E[\dens]$ over $\dens$ then yields the ground-state energy $\Egs$ and density $\densgs$ of the system. 
The density is often expressed as an expansion using atom-centered Gaussian-type basis functions. We follow \cite{zhang2024overcoming, remme2025stable} in applying symmetric Löwdin orthonormalization, yielding an expansion
\begin{equation}
    \rho(\vecr) = \sum_{\mu}p_{\mu}\omega_{\mu}(\vecr)\,,
\end{equation}
with density coefficients $p_\mu$ and orthonormal basis functions $\omega_\mu(\vecr)$.

The energy functional can be split into different physical contributions:
\begin{equation}
    \Etot(\vp) = \Ts(\vp) + \Exc(\vp) + \EH(\vp) + \Eext(\vp),\label{eq:energy_functional}
\end{equation}
where $\Ts$ is the non-interacting kinetic energy, $\Exc$ is the exchange-correlation energy, $\EH$ is the Hartree energy, and $\Eext$ is the interaction energy with an external potential.

In machine-learned OF-DFT~\cite{zhang2024overcoming, remme2025stable, remme2026surrogate}, the energy functional or parts of it are learned by training on labels obtained from Kohn--Sham reference data. Those labels correspond to electron densities at the electronic ground state as well as off-equilibrium densities extracted from intermediate self-consistent field iterations.
During inference, the electron density is initialized using standard methods such as a superposition of atomic densities (SAD). Afterwards, the density is optimized using a gradient-descent scheme where the required gradients of the functional $\grad_{\vp} \Etot$ are obtained by automated differentiation of the neural network. 

\section{\label{sec:methods} Methods}

\subsection{\label{sec:weak_surrogates} Weak surrogate functionals}

Weak surrogate functionals~\cite{remme2026surrogate} lift the requirement of global fidelity to a physical energy functional and instead define a successful functional exclusively through its minimizer in a fixed density optimization procedure. Specifically, a weak surrogate functional $\EWS$ must yield the true ground-state density coefficients $\vpgs$ when minimized from a prescribed initial guess. This allows training using only ground-state density labels, avoiding the need for off-equilibrium densities.

To guarantee convergence during inference, weak surrogates can be trained with the gradient-descent-improvement (GDI) loss. This loss enforces that every optimization step with step size $\lambda$ moves the coefficients closer to the true ground state by at least a predefined contraction factor $0 < \beta < 1$. With $\vp'=\vp-\lambda\grad_\vp\EStot[\vp;\theta]$ and $[x]_+ = \max(0,x)$, the loss \cite{remme2026surrogate} is
\begin{align}\label{eq:gdi}
    \mathcal{L}_{\text{GDI}}
    = \left[\norm*{\vp' - \vpgs} - \beta\norm{\vp - \vpgs}\right]_+ \,.
\end{align}
The contraction factor is chosen such that the GDI loss vanishes for most Kohn--Sham labels from off-equilibrium densities as used in~\cite{remme2025stable}. In this sense, the GDI loss is \emph{compatible} with the physical energy landscape as sampled by these training labels. More details on the choice of $\beta$ and $\lambda$ can be found in Appendix~\ref{app:surrogate_hparams}.

To focus model capacity on the regions actually visited during inference, this loss is combined with an adaptive train-time density optimization scheme (see Figure~\ref{fig:pipeline}). For each molecule $\mathcal{M}$, a cached density coefficient vector $\vp_{\mathcal{M}}$ is maintained and updated during training each time the molecule appears in a batch via a gradient descent step:
\begin{equation}
    \vp_{\mathcal{M}} \rightarrow \vp_{\mathcal{M}} - \gamma\grad_\vp \EStot(\vp_{\mathcal{M}}).
\end{equation}
Note that the step size $\gamma$ may be different from the step size $\lambda$ that is used in the GDI loss. In the original framework, $\gamma$ was kept fixed. Here, we introduce a dynamically computed ``ideal'' learning rate for the cache updates:
\begin{equation}
    \gamma^* = \left[
        \frac{\grad_\vp \EStot(\vp) \cdot (\vp - \vpgs)}
        {\norm*{\grad_\vp \EStot(\vp)}^2}
    \right]_+.
\end{equation}
This rate $\gamma^*$ corresponds to the nonnegative step size that minimizes the Euclidean distance between the updated density coefficients and the ground-state target $\vpgs$ for the currently predicted gradient direction (see Appendix~\ref{app:math_details} for details). Employing this optimal local step size ensures that train-time trajectories progress as efficiently as possible toward the ground state. At the same time, samples for which the predicted gradient would lead to a step away from the ground state are not updated at all, creating a natural training dynamic where difficult-to-learn densities are kept in cache until the model learns to improve the gradient predictions at these points.
\subsection{\label{sec:strong_surrogates} Strong surrogate functionals}

While weak surrogates only ensure convergence to the correct ground-state electron density, strong surrogate functionals are designed to also accurately model the energy at this ground state. From a machine learning perspective, this requires an additional loss term to enforce the true energy at the predicted minimum. However, in the adaptive training scheme described in Section~\ref{sec:weak_surrogates}, the exact ground-state density is never explicitly sampled. The central challenge is therefore how to provide meaningful energy feedback.
To address this, we propose a two-step strategy:
First, we inject a fraction $0\leq q<1$ of exact ground-state samples into each training batch. For these molecules, instead of using their cached density coefficients, we evaluate the model strictly at the true ground-state density and apply $L_1$ penalty losses on both the predicted ground-state energy and its gradients, which should be zero at the ground state: $\mathcal{L}_E = \sum_{\mathcal{M}}\left(|\EStot(\vpgs_{\mathcal{M}}) - \Etot(\vpgs_{\mathcal{M}})|\right)$ and $\mathcal{L}_{\grad E} = \sum_{\mathcal{M}}\left(|\grad_{\vp} \EStot(\vpgs_{\mathcal{M}})|\right)$. Incorporating a fraction of exact ground-state labels also provides the architectural flexibility, in principle, to enforce other ground-state properties, such as nuclear gradients.

\begin{figure}
    \centering
    \includegraphics[width=\linewidth]{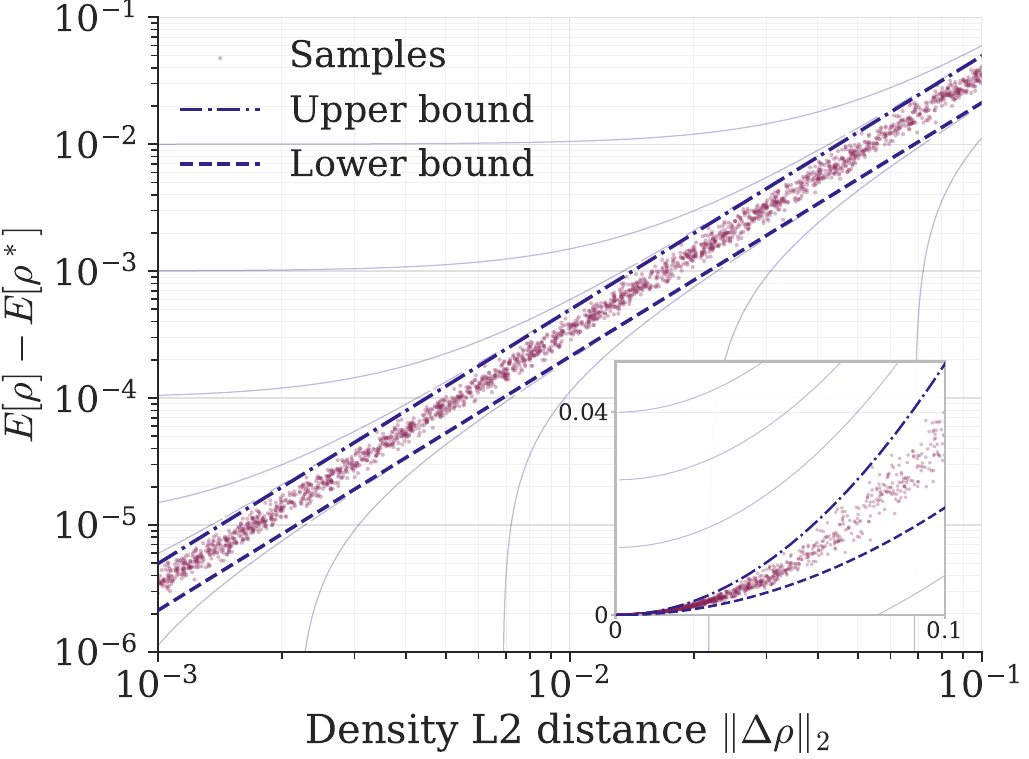}
    \caption{Scatter plot of energy labels from intermediate Kohn--Sham iterations as used in the training set of~\cite{remme2025stable}, with logarithmic scaling on both axes (linear scaling in the inset). The labels show a quadratic dependence on the $L_2$ density difference to the ground state and give rise to quadratic lower and upper bounds that can be used to enforce the energy-bound loss in Equation~\ref{eq:eb_loss}. The fine blue lines indicate isocontours of the loss.}
    \label{fig:empirical_energy_bounds}
\end{figure}

Second, while exact energy labels at arbitrary off-equilibrium densities are not available, we can exploit physical and empirical energy bounds to construct a robust loss function. A natural lower bound is the ground-state energy itself, since it represents the global minimum of the exact functional. However, penalizing predictions only when they fall below the exact ground-state energy yields a very weak learning signal. Instead, we derive tighter empirical lower and upper bounds from off-equilibrium energies and densities, which we take from~\cite{remme2025stable}. As illustrated in Figure~\ref{fig:empirical_energy_bounds}, these labels form a set that, empirically, is bounded above and below by two quadratic forms:
\begin{align}
    E_{\text{min}}(\vp) &= a_{\text{min}} {(\vp - \vpgs)}^T \ovlp (\vp - \vpgs),\\
    E_{\text{max}}(\vp) &= a_{\text{max}} {(\vp - \vpgs)}^T \ovlp (\vp - \vpgs),
\end{align}
where $\ovlp$ is the overlap matrix of the basis functions in the molecule.
Based on this observation, we introduce the following energy-bound loss:
\begin{equation}
    \mathcal{L}_{\text{EB}} =
    \left[E_{\text{min}}(\vp) - \EStot(\vp)\right]_+
    + \left[\EStot(\vp) - E_{\text{max}}(\vp)\right]_+.
    \label{eq:eb_loss}
\end{equation}

The total loss is then given by a weighted sum of all loss components:
\begin{equation}
    \mathcal{L} = \alpha_{1} \mathcal{L}_{\text{GDI}} + \alpha_{2} \mathcal{L}_{\text{EB}} + \alpha_3 \mathcal{L}_E + \alpha_{4} \mathcal{L}_{\grad E}.
\end{equation}

\begin{figure*}[ht]
    \centering
    \begin{tcolorbox}[width=0.48\textwidth, equal height group=molecules, colback=blue!3!white, colframe=blue!5!white, arc=3mm, boxrule=0pt, nobeforeafter, halign=center]
        \textbf{C$_{38}$H$_{43}$F$_{3}$N$_{8}$O$_{3}$} - STRUCTURES25 outlier \\[2ex] %

        \begin{subfigure}{0.48\textwidth}
            \centering
            \tikz[blend mode=multiply] \node[inner sep=0pt, outer sep=0pt] {\includegraphics[angle=90, width=\linewidth]{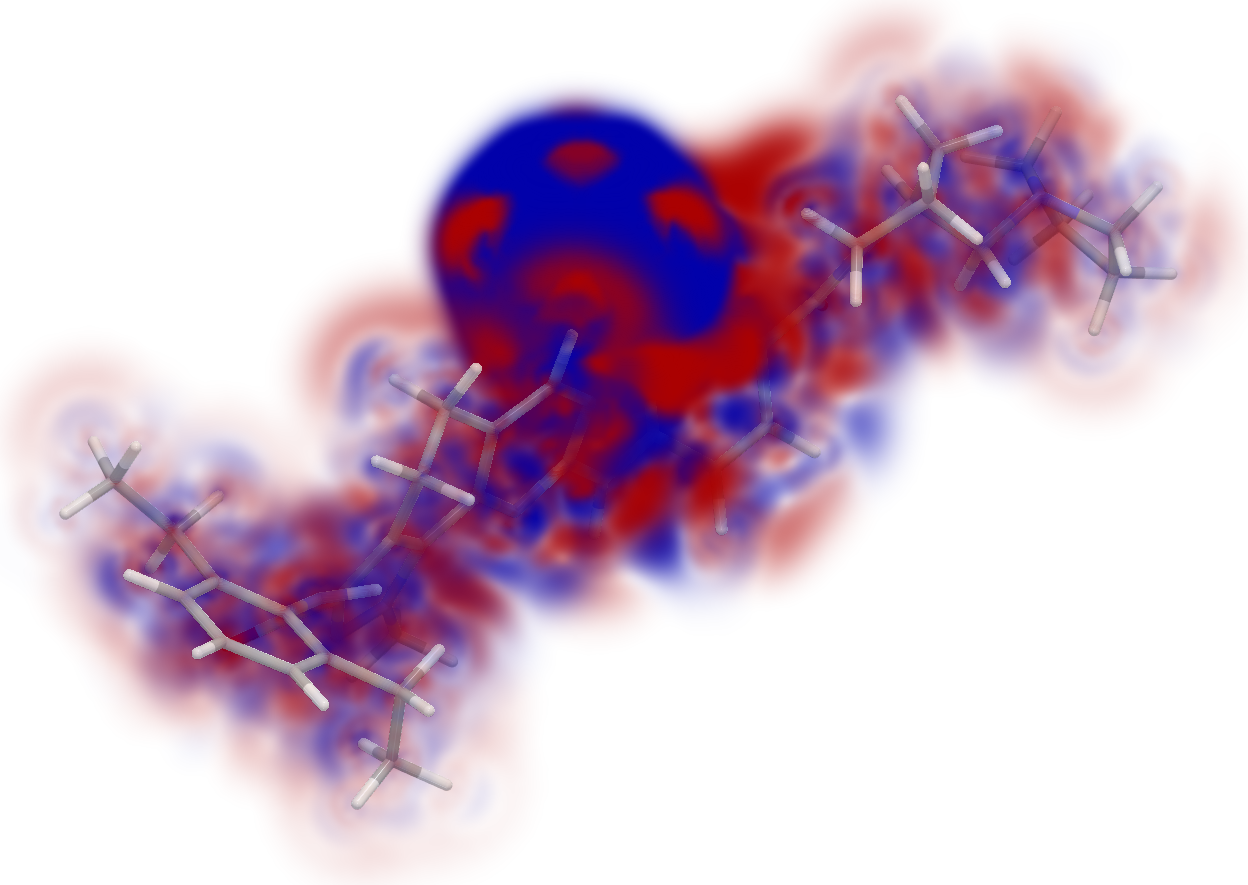}};
            \caption{$\|\Delta \dens\|_2=1.209$}%
            \label{fig:dens_a}
        \end{subfigure}\hfill
        \begin{subfigure}{0.48\textwidth}
            \centering
            \tikz[blend mode=multiply] \node[inner sep=0pt, outer sep=0pt] {\includegraphics[angle=90, width=\linewidth]{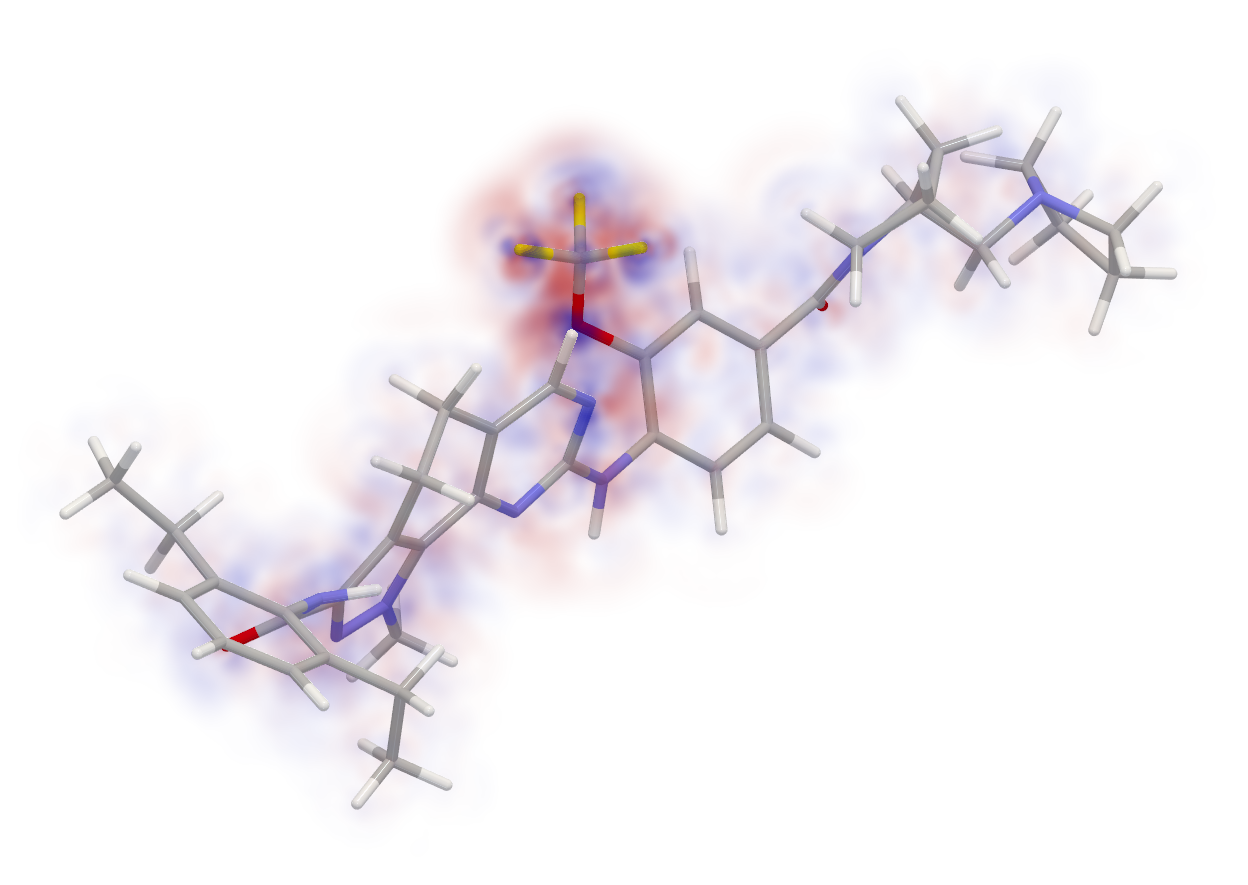}};
            \caption{$\|\Delta \dens\|_2=0.013$}%
            \label{fig:dens_b}
        \end{subfigure}
    \end{tcolorbox}%
    \hfill
    \begin{tcolorbox}[width=0.48\textwidth, equal height group=molecules, colback=blue!3!white, colframe=blue!5!white, arc=3mm, boxrule=0pt, nobeforeafter, halign=center]
        \textbf{C$_{56}$H$_{66}$N$_{8}$O$_{7}$} - Random QMugs molecule\\[2ex] %
        \begin{subfigure}{0.48\textwidth}
            \centering
            \tikz[blend mode=multiply] \node[inner sep=0pt, outer sep=0pt] {\includegraphics[angle=90, width=\linewidth]{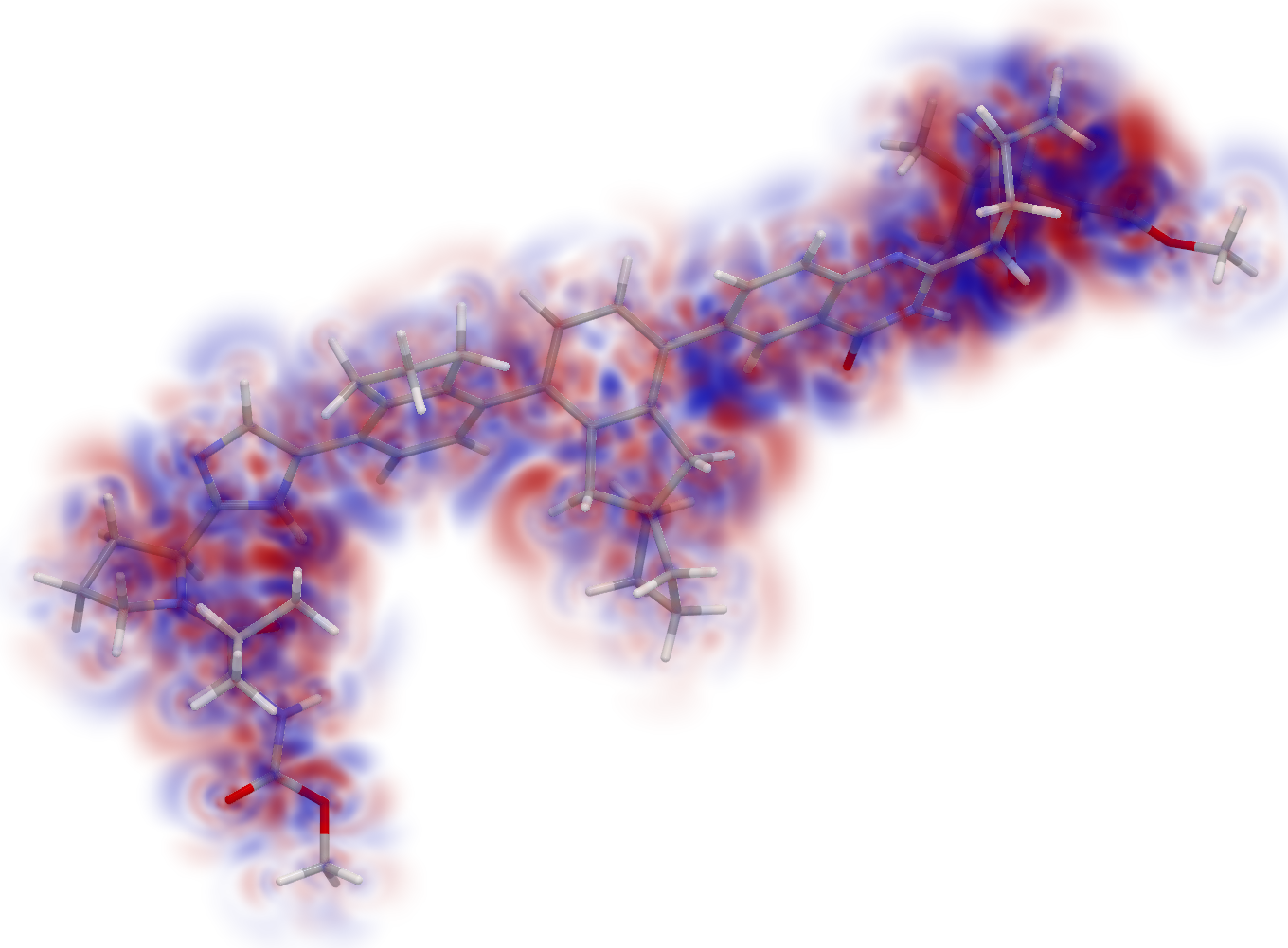}};
            \caption{$\|\Delta \dens\|_2=0.054$}%
            \label{fig:dens_c}
        \end{subfigure}\hfill
        \begin{subfigure}{0.48\textwidth}
            \centering
            \tikz[blend mode=multiply] \node[inner sep=0pt, outer sep=0pt] {\includegraphics[angle=90, width=\linewidth]{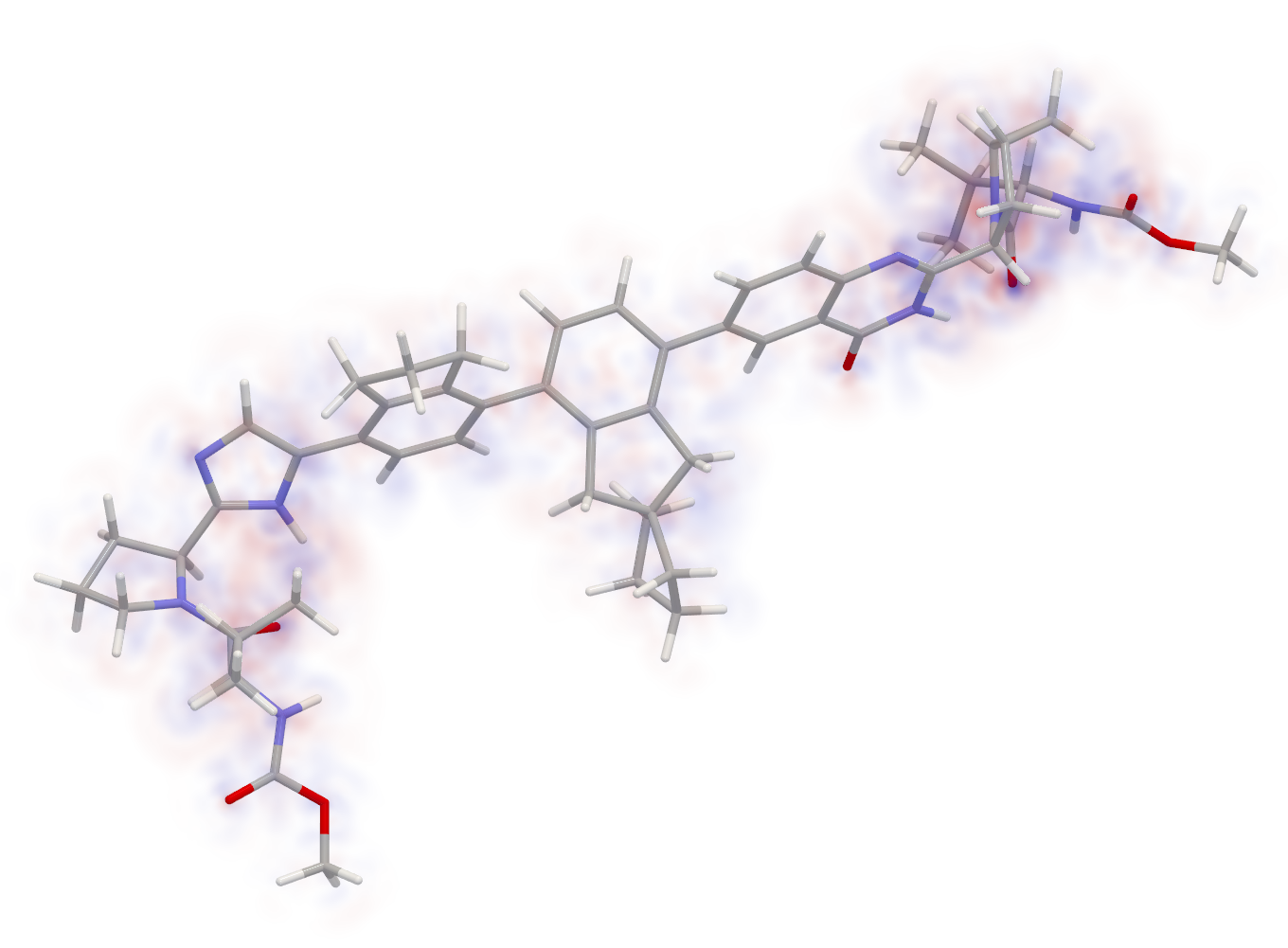}};
            \caption{$\|\Delta \dens\|_2=0.008$}%
            \label{fig:dens_d}
        \end{subfigure}
    \end{tcolorbox}
    \vspace{1ex}
    \caption{Visual comparison of spatially resolved density errors for two molecules from the QMugs dataset. In each panel, the left subfigure shows the density error from the STRUCTURES25 baseline, while the right subfigure displays the error from a strong surrogate functional (this work). The molecule in the left panel corresponds to an outlier of the STRUCTURES25 model, while the molecule in the right panel is a randomly selected example. All plots use the same colorbar.}
    \label{fig:trifluoro_comp}
\end{figure*}

\subsection{\label{sec:txc_training}Training surrogate functionals on $\Etxc$}

The choice of $\Etot$ as energy target is practical, as both the GDI loss and the density cache updates require gradients corresponding to the total energy functional. Nevertheless, there are compelling reasons to use neural networks to model only specific energy components, while computing the remaining contributions using exact physical expressions:
(1) To facilitate generalization across diverse chemical systems, it is advantageous to learn only the universal functional components, isolating them from the system-specific external potential.
(2) Neural network backbones typically apply a radial spatial cutoff, limiting their ability to model long-range interactions. Explicitly evaluating the Hartree energy enables the model to accurately capture the nonlocal electrostatic contributions.
Consequently, learning the sum $\Etxc$ of the non-interacting kinetic energy $\Ts$ and the exchange-correlation energy $\Exc$ represents a physically motivated objective, consistent with prior methodologies~\cite{zhang2024overcoming, remme2025stable}.
In practice this means parameterizing the total energy functional as
\begin{equation}
    \EStot(\vp) = \EStxc(\vp) + \EH(\vp) + \Eext(\vp),
\end{equation}
where $\EStxc$ is the contribution that is learned by the neural network.

\paragraph{Enhancing computational efficiency}

The primary computational bottleneck lies in evaluating the remaining gradient contributions---namely $\grad_{\vp}\EH$ and $\grad_{\vp} \Eext$---at each training step, as these gradients are necessary to perform cache updates and compute the GDI loss. With the convention $\EH(\vp)=\frac{1}{2}\vp^T\coulomb\vp$, the Hartree gradient is given by $\grad_{\vp}\EH = \coulomb \vp$. Computing this term is expensive, requiring either the repeated recalculation of the Coulomb matrix $\coulomb$ for every sample during each epoch, or precomputing and storing $\coulomb$ for the entire dataset, which exhibits poor scaling for large molecular systems.

To enhance the efficiency and scalability of the approach, we propose to use an auxiliary functional $\Eaux(\vp)$ in place of the total energy $\EStot(\vp)$ when computing the GDI loss and performing the cache updates.
This auxiliary functional must preserve the ground-state density of the true energy functional in order for the train-time density optimization scheme to converge to the correct density. A choice that fulfills this criterion while remaining computationally efficient is given by
\begin{equation}
    \Eaux(\vp) = \EStxc(\vp) - \grad_{\vp}\Etxc(\vpgs) \cdot \vp\,,\label{eq:txc_surrogate1}
\end{equation}
where we subtract a linear term parameterized by the gradients of the kinetic and exchange-correlation energies evaluated at the ground state $\vpgs$. We must further ensure that the true ground-state density is also a minimum of $\EStot(\vp)$, which includes the exact Hartree and external energies. Because the gradient of the total energy vanishes at the ground state, we can use Equation~\ref{eq:energy_functional} to express this gradient contribution as
\begin{equation}
    \grad_{\vp}\Etxc(\vpgs) = -\grad_{\vp}\EH(\vpgs) - \grad_{\vp}\Eext(\vpgs).
\end{equation}

Furthermore, since the external energy is a linear functional of the density, we can write $\grad_{\vp}\Eext(\vpgs)\cdot \vp = \Eext(\vp)$. Substituting these relations into Equation~\ref{eq:txc_surrogate1} yields:
\begin{align}
    \Eaux(\vp) &= \EStxc(\vp) + \left(\grad_{\vp}\EH(\vpgs) + \grad_{\vp}\Eext(\vpgs)\right)\cdot \vp,\label{eq:txc_surrogate2}\\
    &= \EStxc(\vp) + \grad_{\vp}\EH(\vpgs)\cdot \vp + \Eext(\vp).\label{eq:txc_surrogate3}
\end{align}
Thus, the auxiliary functional corresponds to linearizing the true Hartree energy around the ground-state density. Using this relation, we can express the residual between the total energy functional and the auxiliary functional as
\begin{align}
    &\EStot(\vp) - \Eaux(\vp) = \EH(\vp) - \grad_{\vp} \EH(\vpgs) \cdot \vp\\
    =& \frac{1}{2}\vp^T\coulomb\vp - {(\vpgs)}^T\coulomb\vp\\
    =& \frac{1}{2}{(\vp - \vpgs)}^T\coulomb(\vp - \vpgs) + \text{const.}
\end{align}
Because the Coulomb matrix $\coulomb$ is positive semi-definite, this residual forms a convex functional. Adding this term to $\Eaux$ therefore preserves the global minimum at $\vpgs$.

As explained above, the GDI and energy-bound losses are designed such that most of the available training labels for off-equilibrium densities have zero loss, in order to impose as few unphysical restrictions as possible. Applying the GDI loss to the auxiliary functional in Equation~\ref{eq:txc_surrogate1} enforces convexity on the sum of the kinetic and exchange-correlation energies---a property not guaranteed for the exact physical functional. 
However, empirical observations suggest that convexity holds for our training data (see Appendix~\ref{app:convexity}). 

\begin{table*}[ht]
\caption{\label{tab:qm9_results} Characterizing in-domain accuracy: Results on QM9 test data. Weak surrogates yield no energies and MLIPs do not predict electron densities. Models marked with a star were trained and evaluated on different QM9 splits including "uncharacterized" molecules, which are usually discarded from the dataset.}
\begin{ruledtabular}
\begin{tabular}{llcccc}
 & Model & Target & $|\Delta E|$ (mHa)& $|\Delta E| / N$ (mHa) & $\|\Delta \dens\|_2 \; (10^{-2})$
\\ \hline
Prior energy functionals& M-OFDFT\textsuperscript{*}\cite{zhang2024overcoming} & $\Etxc$ & 1.37 & 0.088 & 2.7\\
& STRUCTURES25\textsuperscript{*}\cite{remme2025stable} & $\Etxc$&0.64 & 0.038 & 1.4 \\
& Weak surrogate\textsuperscript{*}\cite{remme2026surrogate} & - & - & - & 1.2\\ \hline
MLIPs & eSEN& $\Etot$ & 0.203 & 0.012 & -  \\
& EquiformerV2& $\Etot$ & 0.221 & 0.015 & -  \\
& GotenNet& $\Etot$ & 0.130  & 0.008 & -\\ \hline
Surrogate Functionals & eSEN-SF& $\Etxc$& 0.371 & 0.022 & 0.29 \\
    (this work)& eSEN-SF& $\Etot$& 0.134 & 0.008 & 0.17\\ %
& EquiformerV2-SF& $\Etot$& \textbf{0.125} & \textbf{0.007} & \textbf{0.15}\\ %
\end{tabular}
\end{ruledtabular}
\end{table*}
    
\begin{table*}[ht]
\caption{\label{tab:qmugs_results}%
Characterizing out-of-domain accuracy: Results on large molecules from QMugs. All models were trained on a combined QM9 + QMugs split with up to 15 heavy atoms and evaluated on QMugs molecules up to 100 heavy atoms.}
\begin{ruledtabular}
\begin{tabular}{llcccc}
 & Model & Target & $|\Delta E|$ (mHa)& $|\Delta E| / N$ (mHa) & $\|\Delta \dens\|_2 \; (10^{-2})$
\\ \hline
Prior energy functionals& M-OFDFT \cite{zhang2024overcoming}& $\Etxc$ & 18 & 0.17 & 7.0\\
& STRUCTURES25 \cite{remme2025stable} & $\Etxc$ & 22 & 0.21 & 6.8 \\
& Weak surrogate \cite{remme2026surrogate} & - & - & - & 8.2 \\ \hline
MLIPs & eSEN & $\Etot$& 14.2 & 0.138 & -\\ %
& EquiformerV2 & $\Etot$ & 19.2 & 0.132 & -  \\ %
& GotenNet & $\Etot$ & 17.5 & 0.123 & -  \\ \hline %
Surrogate Functionals & eSEN-SF& $\Etxc$& 8.6 & 0.066 & 1.70\\ %
(this work) & eSEN-SF & $\Etot$& 3.7 & 0.028 & \textbf{1.03}\\ %
& EquiformerV2-SF & $\Etot$& \textbf{3.4} & \textbf{0.026} & 1.08\\ %
\end{tabular}
\end{ruledtabular}  
\end{table*}

\subsection{\label{sec:network_architecture}Network architecture}
Prior work in OF-DFT~\cite{zhang2024overcoming, remme2025stable, remme2026surrogate}, including surrogate functionals for OF-DFT, has modeled the density representation using Graphormer-style architectures~\cite{ying2021transformers}, relying on local canonicalization~\cite{lippmann2025beyond} to express the density coefficients as scalar graph features. SO(3)-equivariant neural networks operating on irreducible tensor features provide a strong alternative for molecular modeling, especially when the target quantities naturally transform under rotations. We therefore evaluate eSEN~\cite{fu2025learning} and EquiformerV2~\cite{liao2024equiformerv2}, two modern equivariant architectures.
Both architectures build on the efficient eSCN formulation~\cite{passaro2023reducing}, which reduces the cost of equivariant convolutions and tensor-product-like interactions. The main architectural difference is the edge aggregation mechanism: eSEN uses edge-conditioned equivariant message passing with unnormalized sum aggregation, whereas EquiformerV2 uses an SO(3)-equivariant multi-head attention block with softmax-normalized edge weights. To feed the density information into these architectures, a new density embedding is created. The coefficients are first rescaled independently in each dimension and subsequently scattered into a fixed-size ``atom-hot'' tensor per atom, as in the Graphormer-style architectures. A single equivariant linear layer is then employed to produce the correct structure of irreducible representations that is needed for both eSEN and EquiformerV2. The resulting embedding is added directly onto the respective embeddings employed in the original architectures. Furthermore, an atomic reference module is used to add element-dependent atomic energy biases to the output of the network. In contrast to~\cite{zhang2024overcoming, remme2025stable}, we do not employ a global energy bias, which we found to degrade the generalization performance to larger molecules.  

\section{\label{sec:experiments} Experiments}

\subsection{Density optimization accuracy}

We train and evaluate our models on molecules from the QM9 dataset~\cite{ramakrishnan2014quantum}. For ground-truth energies and densities, we rely on the same labels utilized in~\cite{remme2025stable}, restricted to ground-state labels, which were generated using the PBE functional~\cite{perdew1996generalized}. Following the methodology established in prior work~\cite{zhang2024overcoming, remme2025stable, remme2026surrogate}, we test the extrapolation capabilities of our approach by training models on a combined pool of QM9 and QMugs~\cite{isert2022qmugs} molecules limited to 15 heavy atoms. We then evaluate the models on a distinct test split of significantly larger QMugs configurations containing up to 100 heavy atoms (refer to Appendix~\ref{app:dataset_details} for full details).

We benchmark our accuracy against established machine-learned orbital-free functionals, namely M-OFDFT~\cite{zhang2024overcoming} and STRUCTURES25~\cite{remme2025stable}, as well as weak surrogate functionals from~\cite{remme2026surrogate}. As additional baselines, we train GotenNet---the current state-of-the-art MLIP on QM9---as well as vanilla eSEN and EquiformerV2 architectures without density embeddings, using only ground-state energy data. Comprehensive model specifications and training hyperparameters are documented in Appendix~\ref{app:model_hparams}. Models trained with the strong surrogate pipeline are denoted with the suffix \mbox{-SF}.

Performance metrics on the QM9 dataset are presented in Table~\ref{tab:qm9_results}. Remarkably, all surrogate variants surpass the existing OF-DFT baselines by a wide margin, demonstrating substantial improvements in both energy predictions and density accuracy. Concurrently, models trained with $\Etot$ targets display notably lower energy errors compared to baseline models trained exclusively on molecular geometries without auxiliary electron density information. 

Identical performance trends persist when evaluating the models on the larger QMugs molecules, as summarized in Table~\ref{tab:qmugs_results}. Notably, the relative performance margin between our strong surrogate functionals and the MLIP baselines widens on this out-of-distribution set. 

While the surrogate functionals trained on $\Etxc$ show higher energy and density errors than their $\Etot$ counterparts, they still outperform all previous OF-DFT baselines trained on the same energy target, demonstrating that the train-time density optimization procedure on the auxiliary functional $\Eaux$ is sufficient to learn a converging functional.

Compared to the OF-DFT baselines, surrogate functionals show density errors that are lower by approximately one order of magnitude. Furthermore, they do not exhibit large outliers as is the case for STRUCTURES25~\cite{remme2025stable}. A visual comparison of density errors for example molecules is shown in Figure~\ref{fig:trifluoro_comp}.

\begin{figure}
    \centering
    \includegraphics[width=\linewidth]{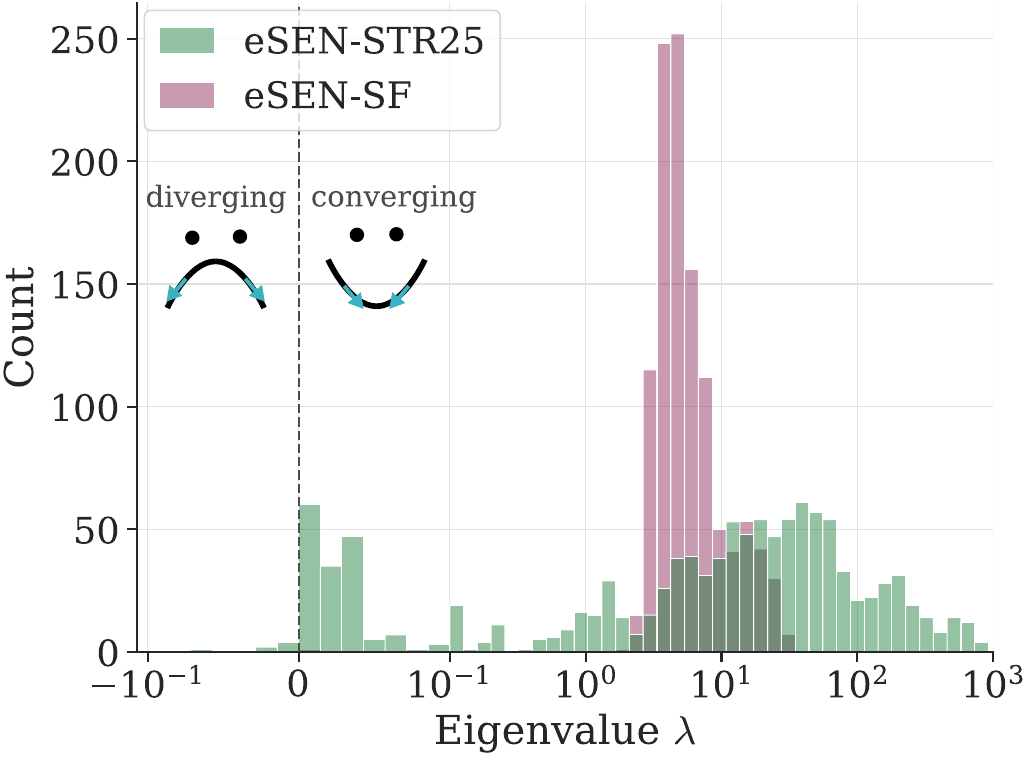}
    \caption{Variational density optimization using prior OF-DFT functionals such as STRUCTURES25 with an eSEN backbone (eSEN-STR25) does not reliably converge, whereas training the same backbone with the surrogate functional pipeline (eSEN-SF) leads to convergent models. The figure depicts the distribution of Hessian eigenvalues at the electronic ground state, where eSEN-STR25 shows a number of negative eigenvalues, corresponding to directions in which the functional is locally concave, while eSEN-SF is locally convex in all directions.}
    \label{fig:energy_landscape}
\end{figure}

It should be noted that all previous machine-learned OF-DFT functionals shown in Tables~\ref{tab:qm9_results} and~\ref{tab:qmugs_results} are based on a Graphormer architecture~\cite{ying2021transformers}, which is less expressive than either eSEN or EquiformerV2 and thus explains part of the performance difference to the surrogate functional. However, we found that retraining STRUCTURES25 with either of the alternative architectures instead of a Graphormer yields a functional that does not converge during density optimization. In Figure~\ref{fig:energy_landscape}, we visualize the distribution of eigenvalues of the Hessian of an eSEN model trained with the STRUCTURES25 training pipeline and the surrogate pipeline, respectively, evaluated at the ground-state density of a random QM9 molecule. The surrogate functional has only positive eigenvalues that all lie within a narrow interval, confirming that the model has indeed learned a convex functional as enforced by the GDI and energy-bound losses. The eSEN-STR25 model, on the other hand, displays a much broader range of Hessian eigenvalues including both much larger and crucially also negative values, leading to a functional that does not converge during density optimization. The fact that we obtain convergence irrespective of the network architecture thus presents a significant advantage of the surrogate-style training.

\subsection{\label{sec:ablations}Ablation experiments}

In order to isolate the contribution of newly introduced components to the strong surrogate training performance, we conduct targeted ablation experiments. We train an EquiformerV2 surrogate functional under two modified constraints: either disabling the energy-bound loss entirely, or replacing the optimally computed ideal learning rate with a fixed cache update learning rate of $10^{-2}$. The resulting evaluations are detailed in Table~\ref{tab:training_ablation_results}. 

\begin{table}[htbp]
\centering
\caption{\label{tab:training_ablation_results} Ablation study demonstrating the impact of omitting key components of the strong surrogate training pipeline.}
\begin{ruledtabular}
\begin{tabular}{cccc}
    \makecell{Energy bound\\loss} & \makecell{Ideal lr \\ (cache update)}  & $|\Delta E|$ (mHa) & $\|\Delta \rho\|_2 \; (10^{-2})$ \\ 
    \hline
    \cmark & \xmark &  0.168 & 0.332 \\ %
    \xmark & \cmark &  0.158  & 0.140 \\ %
    \cmark & \cmark & 0.125 & 0.148 \\ %
\end{tabular}
\end{ruledtabular}
\end{table}

Both ablation scenarios show a degradation in predictive performance, validating the effectiveness of these adaptations within the training pipeline.

\section{\label{sec:conclusion} Discussion and Conclusion}

In this work, we have introduced strong surrogate functionals that achieve state-of-the-art results on PBE energies for both the QM9 and QMugs datasets. These surrogate functionals outperform previous machine-learned orbital-free density functionals by a wide margin. Furthermore, they make the training process more robust, allowing for the use of state-of-the-art neural network backbones while guaranteeing stable convergence during density optimization. 

When evaluated on direct energy prediction, the strong surrogate functionals demonstrate competitive performance relative to standard machine-learned interatomic potentials (MLIPs). While MLIPs can be susceptible to overfitting on datasets of this size, the physical inductive bias embedded within the surrogate framework provides an effective regularization mechanism. This regularization not only improves in-distribution performance but also leads to better generalization on the out-of-distribution QMugs test set.

We further showed that training surrogate functionals on the combined kinetic and exchange-correlation energy components, $\Etxc$, yields a convergent optimization landscape. This modification allows the model to compute long-range electrostatic interactions explicitly through the exact Hartree energy. We attribute the lower accuracy compared to the $\Etot$ models to two main factors: first, the $\Etxc$ term may be a more difficult learning target, and second, the density fitting errors during data generation are absorbed into the $\Etxc$ labels, introducing undesirable artifacts. Moving toward more expressive density representations might thus remedy this issue in the future. 

A current limitation of surrogate functionals, compared to MLIP baselines, is increased computational cost during inference, primarily due to the repeated model evaluations required for density optimization. Future work may alleviate this drawback by designing faster optimizers or using machine-learned direct predictions of the ground-state density coefficients~\cite{zhang2024overcoming}.  

The proposed approach can straightforwardly be extended to molecular labels generated with more accurate XC functionals or even quantum chemistry methods other than Kohn--Sham DFT, as long as both ground-state energies and ground-state densities are available. A promising direction for future work thus lies in training surrogate functionals on larger datasets such as OMol25~\cite{levine2025open} with data generated at the $\omega B97$M-V/def2-TZVPD level of theory, where ground-state electron densities are already available. 

Taken together, these results show that surrogate functionals provide a compelling alternative to MLIPs when electron densities or strong extrapolation are needed, but conventional DFT is computationally prohibitive.

\begin{acknowledgments}
This work is supported by the German Research Foundation (DFG) under Germany's Excellence Strategy EXC-2181/1-390900948 (the Heidelberg STRUCTURES Excellence Cluster) as well as by the Carl-Zeiss-Stiftung via its Wildcard program.
The authors acknowledge support by the state of Baden-W\"urttemberg through bwHPC and the German Research Foundation (DFG) through grant INST 35/1597-1 FUGG.
\end{acknowledgments}

\section*{Data availability statement}

The full training and evaluation code is available at \href{https://github.com/sciai-lab/surrogate-functionals}{https://github.com/sciai-lab/surrogate-functionals}. The data used to train the models is available from Ref. \cite{remme2025stable}.

\appendix

\section{\label{app:math_details} Mathematical details}

\paragraph{Ideal learning rate}

In the following, we provide a derivation for the ideal learning rate $\gamma^*$. Let $\vp$ be the current density coefficients, $\vpgs$ the ground-state coefficients, and $\grad_{\vp} \EStot(\vp)$ the model gradients of the total energy. We want to find $\gamma^*$ such that
\begin{equation}
    \gamma^* = \argmin_{\gamma \geq 0} \|(\vp - \gamma \grad_{\vp} \EStot(\vp)) - \vpgs\|^2_2.
\end{equation}
Taking a derivative w.r.t. $\gamma$ and setting it to zero yields
\begin{align}
    &0\stackrel{!}{=}2 \grad_{\vp}\EStot(\vp) \cdot \left(\left(\vp - \gamma \grad_{\vp} \EStot(\vp)\right) - \vpgs\right)\\
    \iff&0 =  \grad_{\vp}\EStot(\vp) \cdot \left(\vp - \vpgs\right) - \gamma \|\grad_{\vp} \EStot(\vp)\|^2\\
    \iff&\gamma \|\grad_{\vp} \EStot(\vp)\|^2 = \grad_{\vp}\EStot(\vp) \cdot \left(\vp - \vpgs\right)\\
    \iff&\gamma = \frac{\grad_{\vp}\EStot(\vp) \cdot \left(\vp - \vpgs\right)}{\|\grad_{\vp} \EStot(\vp)\|^2}
\end{align}
The nonnegativity constraint therefore gives the projected optimum used in the cache update:
\begin{equation}
    \gamma^* = \left[
        \frac{\grad_{\vp}\EStot(\vp) \cdot \left(\vp - \vpgs\right)}
        {\|\grad_{\vp} \EStot(\vp)\|^2}
    \right]_+.
\end{equation}

\section{\label{app:convexity}Empirical convexity of the energy functional}

\begin{figure}[htbp]
    \centering
    \includegraphics[width=\linewidth]{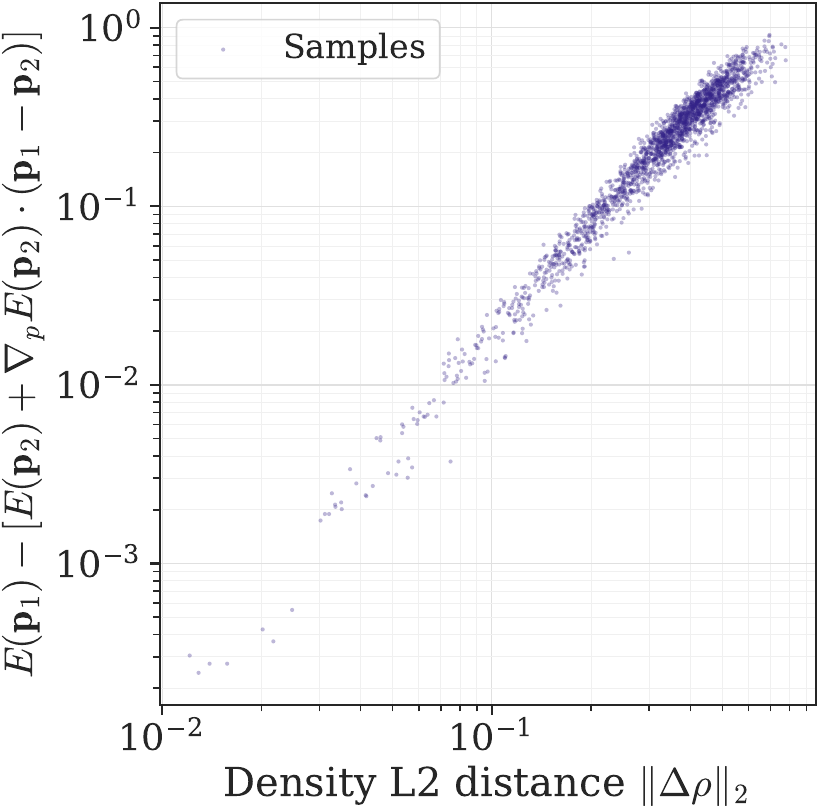}
    \caption{Energy residuals between $\Etxc$ and its linearization at off-equilibrium density labels for a random subset of QM9 molecules. All values are positive, indicating that the $\Etxc$ labels from~\cite{remme2025stable} are consistent with a convex functional.}
    \label{fig:convexity_e_txc}
\end{figure}

One design philosophy of surrogate functionals and the applied losses is to impose as few unphysical constraints as possible. As discussed in Section~\ref{sec:txc_training}, the GDI loss enforces convexity on the energy target the model is trained on. While this convexity holds for the total energy functional $\Etot$ in most cases, it is not strictly guaranteed for the $\Etxc$ functional.\footnote{Training an $\EStxc$ surrogate functional with the objective described in Section~\ref{sec:txc_training} technically enforces convexity of $\Eaux$. However, since the difference to $\EStxc$ is a functional that is linear in $\vp$, this imposes the same convexity constraints also on $\EStxc$.} To investigate how much of a problem this is in practice, we evaluate the empirical convexity of off-equilibrium $\Etxc$ labels from~\cite{remme2025stable} for a random subset of QM9 molecules. For each molecule, we sample two labels $(\vp_1, E_{\text{txc},1}), (\vp_2,  E_{\text{txc},2})$ and compute the residual between $\Etxc(\vp_1)$ and the energy at $\vp_1$ obtained by linearizing $\Etxc$ at $\vp_2$:
\begin{equation}
    \Etxc(\vp_1) - \left[\Etxc(\vp_2) + \grad_{\vp}\Etxc(\vp_2)\cdot (\vp_1 - \vp_2)\right].
\end{equation}
For a convex functional, this residual should be positive for all pairs of labels. We plot the distribution of this residual in Figure~\ref{fig:convexity_e_txc} and find that indeed all values are positive. This suggests that the convexity we impose on $\Etxc$ is at least compatible with our PBE training data.

\section{Training details}

\subsection{\label{app:dataset_details}Details on datasets}

We evaluate our models on the QM9~\cite{ramakrishnan2014quantum} and QMugs~\cite{isert2022qmugs} datasets. As ground-truth targets, we rely on ground-state energies and electron densities computed at the PBE level of theory~\cite{perdew1996generalized}, identical to the data used by Remme et al.~\cite{remme2025stable}. For the density labels, an even-tempered basis with $\beta=2.5$ is fitted to the density in the Kohn--Sham orbital basis.

The QM9 dataset comprises small organic molecules consisting of C, H, O, N, and F atoms, containing up to 9 heavy atoms. In contrast to other OF-DFT baselines~\cite{zhang2024overcoming, remme2025stable,remme2026surrogate}, which trained and evaluated their models on a split that included molecules typically flagged as uncharacterized, we opt for the more common QM9 splitting strategy where these uncharacterized molecules are discarded.

To evaluate out-of-distribution generalization to more expansive molecular structures, we train on a superset of the QM9 dataset as well as molecules from the QMugs dataset with up to 15 heavy atoms. The models are then tested on a separate, distinct test set of 850 large QMugs molecules containing up to 100 heavy atoms.

\subsection{\label{app:model_hparams}Model hyperparameters}

For a fair comparison between the surrogate functionals and the direct energy prediction models, we use the same hyperparameters for the eSEN and EquiformerV2 backbones in both cases. The hyperparameters are listed in Tables~\ref{tab:esen_hparams} and~\ref{tab:equiformer_hparams}, respectively.
The values for $l_\text{max}$ were chosen as 4 since this corresponds to the highest angular momentum in our density basis. We also found that choosing a higher cutoff radius than 6 Bohr could strongly degrade the performance when looking at the generalization to larger molecules. 

\begin{table}[h]
\caption{\label{tab:esen_hparams}
Hyperparameters of the eSEN network used for both the direct energy prediction
experiments as well as the surrogate functionals.
}
\centering
\begin{tabular}{@{} l @{\hspace{1.5cm}} l @{}}
\hline\hline
Hyperparameter & Value \\
\hline
\# sphere channels & 32\\
$l_{\text{max}}$ & 4\\
$m_{\text{max}}$ & 4\\
max neighbors & 300\\
cutoff radius & 6 Bohr\\
\# edge channels & 32\\
distance function & gaussian\\
\# distance basis & 128\\
\# layers & 4\\
\# hidden channels & 64\\
normalization type & rms\_norm\_sh\\
activation type & gate\\
ff\_type & spectral\\
\hline
\textbf{\# Parameters} & \textbf{2.7 M}\\
\hline\hline
\end{tabular}
\end{table}

\begin{table}[h]
\caption{\label{tab:equiformer_hparams}%
Hyperparameters of the EquiformerV2 used for both the direct energy prediction experiments as well as the surrogate functionals.
}
\begin{tabular}{@{} l @{\hspace{1.5cm}} l @{}}
\hline\hline
Hyperparameter & Value \\ \hline 
\# sphere channels & 64\\
$l_{\text{max}}$ & 4\\
$m_{\text{max}}$ & 4\\
max neighbors & 300\\
cutoff radius & 6 Bohr\\
\# edge channels & 64\\
distance function & gaussian\\
\# distance basis & 128\\
\# layers & 4\\
\# heads & 4 \\
\# attn\_hidden\_channels & 32\\
\# attn\_alpha\_channels & 32\\
\# attn\_value\_channels & 8\\
\# ffn\_hidden\_channels & 64\\
normalization type & layer\_norm\_sh\\\hline
\textbf{\# Parameters} & \textbf{2.8 M} \\
\hline\hline
\end{tabular}
\end{table}

\subsection{\label{app:surrogate_hparams}Surrogate training details}

\begin{figure}[h]
    \centering
    \includegraphics[width=\linewidth]{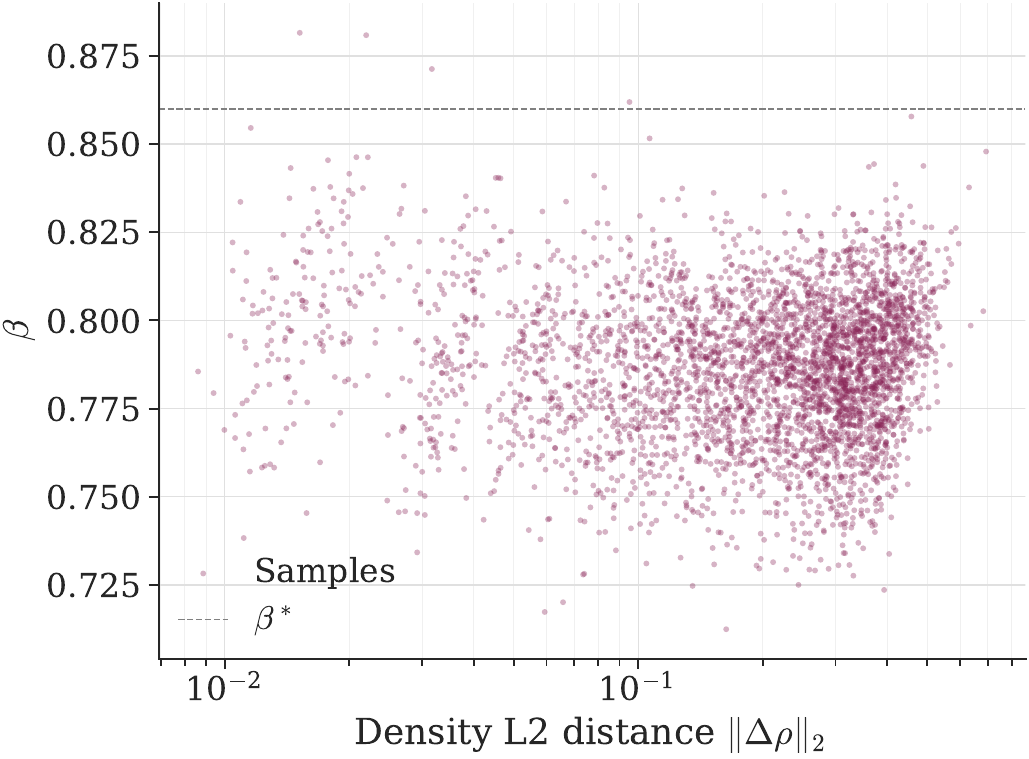}
    \caption{Contraction factors for the gradient labels of 1000 random samples of the QM9 dataset. The value of $\beta$ is chosen as a rough upper bound (ignoring some outliers) and shown as a horizontal line.}
    \label{fig:beta_choice}    
\end{figure}

As described in Section~\ref{sec:weak_surrogates}, we want to choose the hyperparameters of the GDI loss $\lambda$ and $\beta$ such that the loss imposes as few unphysical constraints as possible. To this end, we look at the contraction factors $\{\beta_i(\lambda)\}_{i=1}^{N}$ of physical gradient labels from intermediate Kohn--Sham iterations for different step sizes $\lambda$ and choose the parameters as
\begin{align}
    \lambda^* &= \arg\min_\lambda \max_{i \in \{1, \dots, N\}} \beta_i(\lambda) \label{eq:lambda_opt}\\
    \beta^* &= \max_{i \in \{1, \dots, N\}} \beta_i(\lambda^*) \label{eq:beta_opt}
\end{align}
The optimization in Equations~\ref{eq:lambda_opt} and~\ref{eq:beta_opt} is only performed in an approximate manner by looking at a discrete set of step sizes $\lambda$ and by ignoring some outliers in the distribution of contraction factors. For the choice of $\lambda^* = 0.04$, the distribution of contraction factors as well as $\beta^*$ are visualized in Figure~\ref{fig:beta_choice}.

We list the remaining surrogate-specific hyperparameters as well as general training settings for eSEN-SF and EquiformerV2-SF in Table~\ref{tab:surrogate_hparams}.

\begin{table}[h]
\caption{\label{tab:surrogate_hparams}%
Hyperparameters for surrogate training.
}
\begin{tabular}{lcc}
\hline\hline
Hyperparameter & $\Etot$ training & $\Etxc$ training \\ \hline 
  fraction of surrogate labels & 0.5 & "\\
  cache reset probability & 0.02 & "\\
  cache step size & adaptive & "\\
  GDI step size & 0.04 & adaptive\\
  $\beta$ & 0.86 & 0.925 \\
  GDI loss weight & 20 & "\\ 
  EB loss weight & 1 & " \\
  energy loss weight & 1 & " \\
  gradient loss weight & 1 & " \\ \hline
  batch size & 128 & "        \\
  optimizer & Adam & " \\
  learning rate schedule & cosine & " \\
  maximum learning rate & $1 \cdot 10^{-3}$ & "\\
  \# epochs & 500 & "\\
  \hline\hline  
\end{tabular}
\end{table}

\subsection{\label{app:direct_energy_training}Direct energy training details}

For training the MLIP models on the task of direct energy prediction, we use the settings reported in Table~\ref{tab:esen_training_hparams}. In contrast to the training of the surrogate functionals, we include an exponential moving average of the model parameters and use weight decay along with the AdamW optimizer to reduce overfitting. We train until convergence of the validation loss.

\begin{table}[h]
\caption{\label{tab:esen_training_hparams}%
Hyperparameters for training the eSEN and EquiformerV2 model on the task of direct energy prediction.
}
\begin{tabular}{@{} l @{\hspace{1.5cm}} l @{}}
\hline\hline
Hyperparameter & Value \\ \hline 
batch size & 128 \\
optimizer & AdamW \\
learning rate schedule & cosine \\
maximum learning rate & $4 \cdot 10^{-4}$\\
\# epochs & 500 \\
weight decay & $1 \cdot 10^{-3}$\\
Model EMA decay & $0.999$\\
\hline\hline
\end{tabular}
\end{table}

\subsection{\label{app:additional_results}STRUCTURES25 training without global energy bias}
As mentioned in Section~\ref{sec:network_architecture}, we do not use a global energy bias for our models. For a fairer comparison with the non-surrogate ML functionals, we retrained STRUCTURES25 on the combined QM9 and QMugs split without a global energy bias and report the inference results on the large test molecules in Table~\ref{tab:str25_qmugs_results}.

\begin{table}[h]
\caption{\label{tab:str25_qmugs_results}%
STRUCTURES25 results on large molecules from QMugs when trained without a global energy bias.}
\begin{tabular}{@{} l@{\hspace{0.5cm}}l@{\hspace{0.5cm}}l @{}}
\hline\hline
 $|\Delta E|$ (mHa)& $|\Delta E| / N$ (mHa) & $\|\Delta \dens\|_2 \; (10^{-2})$
\\ \hline
8.6 & 0.09 & 8.2  \\
\hline \hline
\end{tabular}
\end{table}

The energy predictions significantly improve compared to the model containing a global bias, while still being worse than the predictions of the surrogate functionals. 

\bibliography{references}%

\end{document}